\pdfoutput=1
\documentclass[conference]{IEEEtran}
\IEEEoverridecommandlockouts
\usepackage{cite}
\usepackage{amsmath,amssymb,amsfonts}
\usepackage{graphicx}
\usepackage{textcomp}
\usepackage{xcolor}
\usepackage{booktabs}

\definecolor{eprop}{HTML}{2a78d6}
\definecolor{stpx}{HTML}{eb6834}
\definecolor{learn}{HTML}{138a62}
\definecolor{epropd}{HTML}{17498d}
\definecolor{stpxd}{HTML}{9a3a0e}
\newcommand{\co}[1]{\textcolor{eprop}{#1}}
\newcommand{\ct}[1]{\textcolor{stpx}{#1}}
\newcommand{\cl}[1]{\textcolor{learn}{#1}}
\newcommand{\cod}[1]{\textcolor{epropd}{#1}}
\newcommand{\ctd}[1]{\textcolor{stpxd}{#1}}
\newcommand{\dd}{\mathrm{d}}

\begin{document}

\title{Local credit assignment through short-term synaptic states in spiking networks}

\author{\IEEEauthorblockN{Mauricio A. Diaz\textsuperscript{1,2,*}\hspace{2.4em}Janina Hesse\textsuperscript{1,2,3,*}}
\IEEEauthorblockA{\textsuperscript{1}\textit{Computational Resilience Group, Leibniz Institute for Resilience Research (LIR), Mainz, Rhineland-Palatinate, Germany}\\
\textsuperscript{2}\textit{Institute for Quantitative and Computational Biosciences (IQCB), Johannes Gutenberg University,}\\ \textit{Mainz, Rhineland-Palatinate, Germany}\\
\textsuperscript{3}\textit{University Medical Center of the Johannes Gutenberg University, Mainz, Rhineland-Palatinate, Germany}\\
\textsuperscript{*}mauricio.a-diaz@lir-mainz.de, janina.hesse@lir-mainz.de}}

\maketitle

\begin{abstract}
Short-term synaptic plasticity (STP) gives spiking networks a memory, because the transient state of a synapse retains a trace of past spikes. To use this memory for a task with delayed outcomes, learning must propagate errors through these synaptic states. The gradient computed by backpropagation through time (BPTT) includes this path, whereas eligibility propagation (e-prop), a local approximation of it, omits it. E-prop can learn to read a synaptic memory, through the synaptic release in its eligibility trace, but not to write it. From an exact factorisation of the gradient, we derive STP-extended e-prop (STPx e-prop), which preserves the sensitivity of outgoing synaptic states exactly, as a second eligibility trace, and approximates the associated credit as e-prop does, with random feedback of which the new trace uses only the sign. The resulting rule combines locally computed traces with broadcast output errors and requires no weight transport. On delayed association without an initial cue preference, STPx e-prop achieves a test accuracy of 0.95 (BPTT 0.96), whereas e-prop and BPTT without this path remain at chance. On delayed XOR, it achieves 0.96 (BPTT 0.99), where e-prop also remains at chance, and on evidence accumulation it outperforms e-prop by 0.12--0.13. These results show that spiking networks can learn to write task-relevant information into short-term synaptic memory using local signals, as neuromorphic on-chip learning requires.
\end{abstract}

\begin{IEEEkeywords}
spiking neural networks, short-term synaptic plasticity, e-prop, local learning, neuromorphic computing
\end{IEEEkeywords}

\section{Introduction}\label{sec-intro}

Learning from delayed behavioural outcomes requires assigning credit to earlier neuronal activity. Synaptic eligibility traces, slowly decaying biophysical processes in activated synapses~\cite{gerstner2018-eligibility}, provide a potential solution by retaining activity-dependent information that can interact with a later teaching or modulatory signal. Short-term synaptic plasticity (STP) introduces a further temporal dependency: facilitation and depression allow past spikes to shape subsequent synaptic transmission through transient presynaptic states~\cite{tsodyks1998-neural}. These states have been proposed as a substrate for working memory without sustained, cue-selective firing~\cite{mongillo2008-synaptic}. Learning through such synaptic memories, however, requires assigning credit not only to the spikes that read out stored information, but also to the earlier spikes that established the states of outgoing synapses.

Recurrent networks trained with backpropagation through time (BPTT) can learn to retain cue information in synaptic states~\cite{masse2019-circuit,kozachkov2022-robust}, but standard BPTT requires storing intermediate states and propagating gradients backward through the trial~\cite{werbos1990-backpropagation}. Eligibility propagation (e-prop) instead combines forward-running local eligibility traces with approximate learning signals~\cite{Bellec2020-yh}. Presynaptic STP introduces a further path: STP facilitates or depresses synaptic transmission transiently following spiking activity (Fig.~\ref{fig-model}a), i.e. an incoming weight changes a neuron's spikes, which modify its outgoing synaptic states and affect later transmission to downstream neurons. The exact e-prop factorisation contains this path in its learning signals, but e-prop is used as a local rule, and the direct output feedback that makes it local omits the corresponding credit. Networks that learn to use a synaptic memory are therefore trained with BPTT, along a credit path for which no local rule is known, and whether a circuit could learn to use such a memory without backpropagation remains open.

Here, we introduce STP-extended e-prop (STPx e-prop), which augments the eligibility trace of each synapse with a vector tracking the sensitivity of the neuron's outgoing Tsodyks--Markram synaptic state to the incoming weight, and adds one feedback weight per neuron (Fig.~\ref{fig-model}b). Starting from the exact gradient, we derive the contribution through this state and introduce one approximation beyond those of e-prop: replacing the downstream neuron's surrogate-derivative gate with the sender's gate. The resulting rule retains a memory-writing path without a backward pass through the trial. In its final form, STPx e-prop uses random output-error feedback, like random e-prop~\cite{Bellec2020-yh}, and replaces the remaining non-local feedback weight with the sign of the neuron's random feedback weight and a fixed gain; it is therefore fully local and requires no weight transport.

We test the rule on three tasks. On delayed association from an initialisation without cue preference, which isolates the writing of the memory, STPx e-prop performs close to BPTT in networks of 20 to 200 neurons. E-prop remains at chance, and similarly two other controls: a rule that receives the additional feedback weight but lacks the trace of the synaptic state, and BPTT with the path through the synaptic state removed (Sections \ref{sec-write} and \ref{sec-needs}). E-prop with the release in its eligibility trace learns where a cue preference is present at initialisation, and not with the presynaptic spike in its place. Making the rule local costs no accuracy: neither random feedback, nor its sign in place of the feedback weight, nor the fixed gain reduces it (Section~\ref{sec-needs}). On delayed XOR, in which no single input weight correlates with the target, STPx e-prop reaches 0.96 with facilitating synapses, where e-prop remains near chance, and STP in half of the neurons suffices (Section~\ref{sec-xor}). The rule also learns evidence accumulation in networks without adaptive neurons (Section~\ref{sec-ea}).

Online approximations of real-time recurrent learning~\cite{williams1989-learning} differ in which parts of the influence of a weight on the network state they retain~\cite{marschall2019-unified}: e-prop~\cite{Bellec2020-yh} and random-feedback local online learning~\cite{Murray2019-oe} retain only the influence on the neuron that receives the weight, the multidigraph learning rule~\cite{Liu2021-it} and ModProp~\cite{liu2022-biologically} restore the influence of a spike on its targets, and STPx e-prop restores the influence that a spike exerts through the state of the outgoing synapses. STPx e-prop avoids weight transport with random feedback weights, with which the weights of the network align during learning~\cite{lillicrap2016-random}. Short-term plasticity has also been used to route credit, by separating a feedback signal carried by bursts from the feedforward signal~\cite{payeur2021-burst}; there, STP transmits the teaching signal, whereas here it holds the memory through which credit is assigned. Most closely related, Shimizu and Toyoizumi derive eligibility-based gradients for postsynaptic efficacy and presynaptic release probability under a Fisher-information objective~\cite{shimizu2026-reshaping}; we instead address temporal credit assignment through synaptic memory. The code and simulation data are available at https://github.com/mjadiaz/stpx-eprop and https://doi.org/10.5281/zenodo.24604653.

\section{Model and tasks}

\subsection{Network}\label{sec-network}

\begin{figure*}[t]
\centering
\begin{minipage}[b]{280pt}\centering
\raisebox{-0.5\height}{\includegraphics[width=150pt]{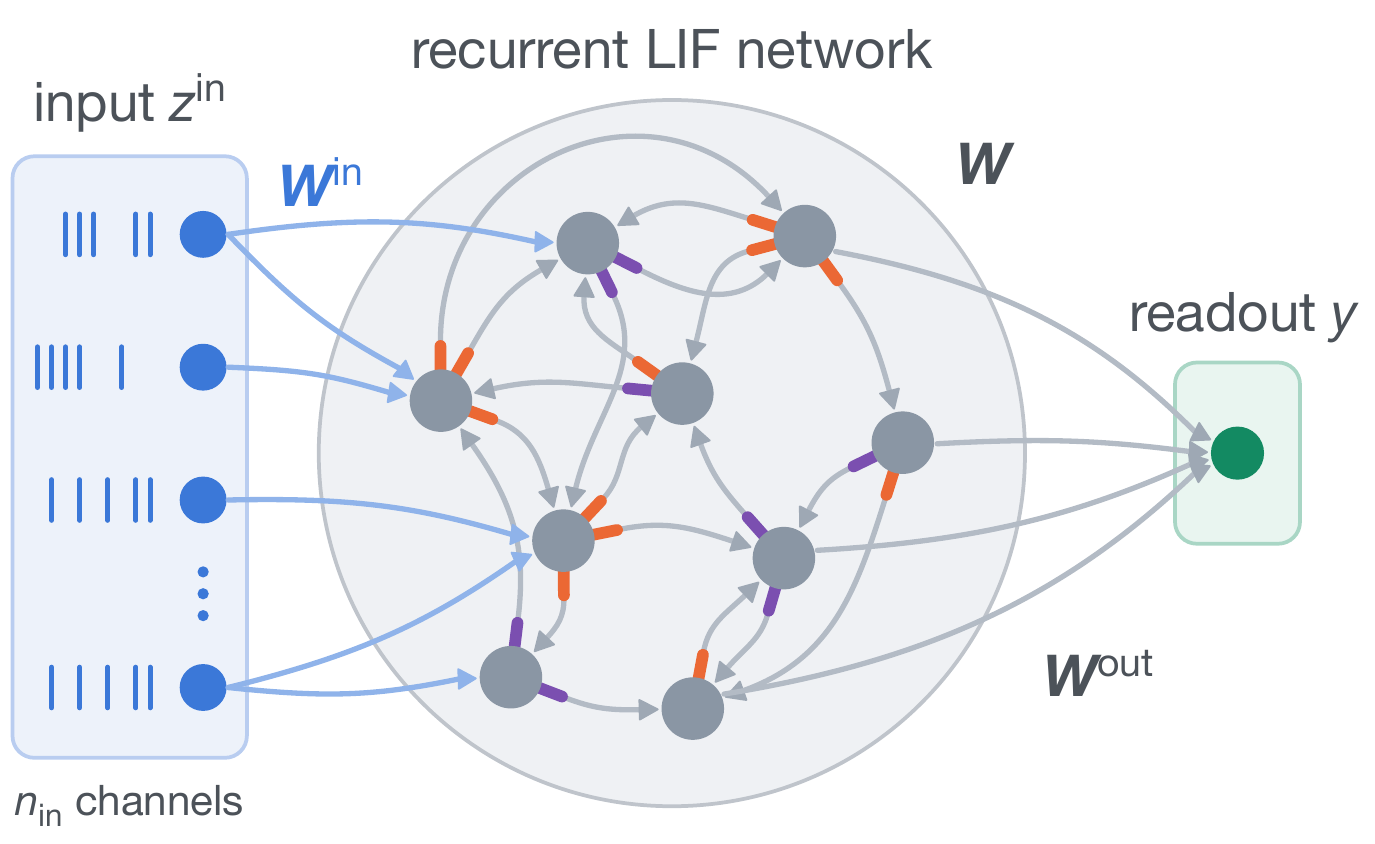}}\hspace{6pt}\raisebox{-0.5\height}{\includegraphics[width=118pt]{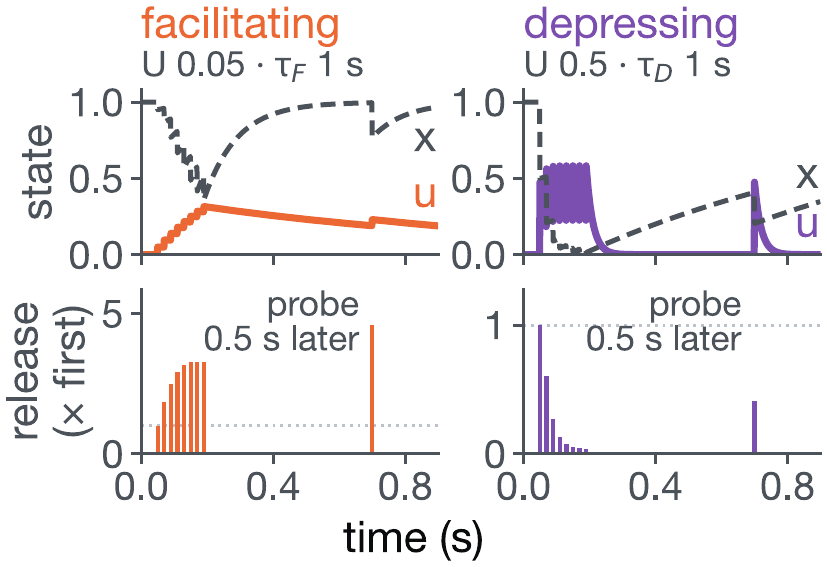}}\\[3pt]
{\footnotesize (a) network and TM synapses}
\end{minipage}\hspace{10pt}\begin{minipage}[b]{222pt}\centering
\includegraphics[width=222pt]{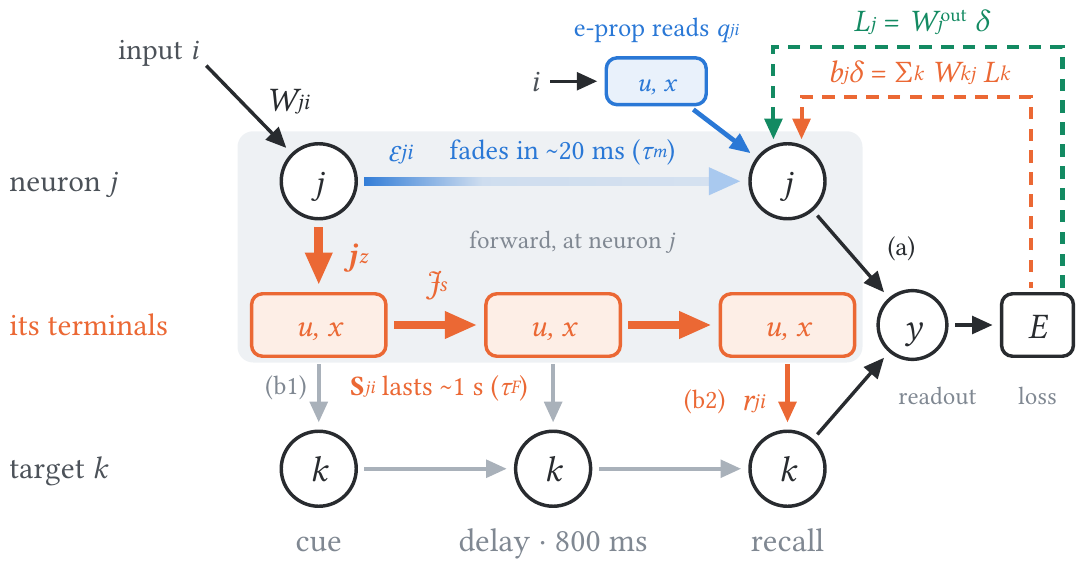}\\[1pt]
{\scriptsize \textcolor[HTML]{a9b1ba}{\rule[0.3em]{1.5em}{0.8pt}}\hspace{0.3em}BPTT only: (b1) and longer paths\hspace{0.7em}\textcolor[HTML]{2a78d6}{\rule[0.3em]{1.5em}{1.6pt}}\hspace{0.3em}\co{e-prop}\hspace{0.7em}\textcolor[HTML]{eb6834}{\rule[0.3em]{1.5em}{1.6pt}}\hspace{0.3em}\ct{STPx e-prop adds}}\\[3pt]
{\footnotesize (b) credit paths of a weight}
\end{minipage}
\caption{Model and credit paths of a weight. (a) Left: recurrent network of LIF neurons with input spikes $z^{\mathrm{in}}$ (weights $W^{\mathrm{in}}$), recurrent weights $W$ and a readout $y$ (weights $W^{\mathrm{out}}$). Each recurrent connection leaves its sender through a facilitating (orange) or depressing (purple) terminal; by default, all recurrent synapses are facilitating. Right: utilisation $u$ (solid), available resources $x$ (dashed), and release relative to the first spike (bars) of a facilitating ($U = 0.05$, $\tau_{F} = 1$ s) and a depressing ($U = 0.5$, $\tau_{D} = 1$ s) Tsodyks--Markram synapse, during a burst of eight spikes at 50 Hz and a probe spike 0.5 s later. (b) Credit paths from a weight $W_{j i}$ to the loss across one trial. Rows show neuron $j$, the state $(u, x)$ of its outgoing synapses (terminals), and a target $k$; colours indicate which gradient follows each path, and (a), (b1) and (b2) mark the three paths of Eq.~\eqref{eq-exact}. Dashed arrows: learning signals $L_{j} = W_{j}^{\mathrm{out}} \delta$ and $b_{j} \delta = \sum_{k} W_{k j} L_{k}$. Both STPx rules follow the orange path.}
\label{fig-model}
\end{figure*}

We simulate $n$ leaky integrate-and-fire neurons with time step $\Delta t = 10$ ms (Fig.~\ref{fig-model}a):
\begin{equation}
\begin{aligned}
  v_{j}^{t} &= \alpha v_{j}^{t-1} + (1 - \alpha) (\sum_{i=1}^{n} W_{j i} q_{j i}^{t} + \sum_{a=1}^{n_{\mathrm{in}}} W_{j a}^{\mathrm{in}} z_{a}^{\mathrm{in}, t}) - \theta z_{j}^{t-1}, \\
  z_{j}^{t} &= H(v_{j}^{t} - \theta),
\end{aligned}
\label{eq-lif}
\end{equation}
where $\alpha = e^{-\Delta t/\tau_{m}}$, $\tau_{m} = 20$ ms, $\theta$ is the firing threshold, and $H$ is the Heaviside step function. Spikes trigger a subtractive reset, $- \theta z_{j}^{t-1}$. The second sum represents $n_{\mathrm{in}}$ external spike inputs $z_{a}^{\mathrm{in}, t}$; input synapses $W^{\mathrm{in}}$ are static. The first sum represents recurrent synaptic release, where $q_{j i}^{t}$ is the release from neuron $i$ onto neuron $j$; recurrent synapses link every pair of distinct neurons with a baseline weight $W$, and the release follows Tsodyks--Markram (TM) dynamics, which depend on the available resources $x$ (depression variable) and the utilisation $u$ (facilitation variable)~\cite{tsodyks1998-neural,mongillo2008-synaptic}:
\begin{equation}
\begin{aligned}
  u^{+} & = u_{j}^{t} + U (1 - u_{j}^{t}) z, \quad q_{k j}^{t} = A_{0} u^{+} x_{j}^{t} z, \\
  u_{j}^{t+1} & = d_{F} u^{+}, \quad x_{j}^{t+1} = 1 - d_{D} (1 - x_{j}^{t} + u^{+} x_{j}^{t} z),
\end{aligned}
\label{eq-tm}
\end{equation}
where $d_{F} = e^{-\Delta t / \tau_{F}}$ and $d_{D} = e^{-\Delta t / \tau_{D}}$. At step $t$ the synapse processes the spike of the previous step, $z = z_{j}^{t-1}$: it releases $q_{k j}^{t}$ from the state $\boldsymbol{s}_{j}^{t} = (u_{j}^{t}, x_{j}^{t})$ and produces $\boldsymbol{s}_{j}^{t+1}$.

All outgoing synapses of neuron $j$ share parameters and a single state $\boldsymbol{s}_{j} = (u_{j}, x_{j})$. The default synapse is facilitating, with $U = 0.05$, $\tau_{F} = 1$ s, and $\tau_{D} = 100$ ms. $A_{0}$ normalises the mean release to 1 per spike for Poisson firing at 10 Hz ($A_{0} = 3.7$ for the default synapse). Unless stated otherwise, all recurrent synapses are facilitating, as are the synapses within a subnetwork of pyramidal cells in ferret prefrontal cortex~\cite{wang2006-heterogeneity}. Networks with depressing synapses use $U = 0.5$, $\tau_{D} = 1$ s and $\tau_{F} = 20$ ms.

The readout is a weighted sum of low-pass-filtered spikes, with loss evaluated over the recall window:
\begin{equation}
\begin{aligned}
  f_{j}^{t} & = \kappa f_{j}^{t-1} + (1 - \kappa) z_{j}^{t}, \quad y^{t} = \sum_{j} W_{j}^{\mathrm{out}} f_{j}^{t}, \\
  E & = \frac{1}{2M} \sum_{t} m^{t} (y^{t} - y^{* t})^{2},
\end{aligned}
\label{eq-readout}
\end{equation}
where $\kappa = e^{-\Delta t / (50\,\mathrm{ms})}$, $m^{t}$ indicates the recall window, $M = \sum_{t} m^{t}$, and $\cl{\delta^{t}} = \partial E / \partial y^{t} = m^{t} (y^{t} - y^{* t}) / M$ is the output error. 

In networks with STP in half of the neurons, all outgoing synapses of the other neurons are static. In control networks with adaptive neurons, half of the neurons have a spike-triggered adaptive threshold~\cite{Bellec2020-yh} and all synapses are static.

\subsection{Tasks}\label{sec-tasks}

\emph{Delayed association.} A 50 ms cue at 80 Hz on input group A or B (four channels each) is followed by an 800 ms delay and a 50 ms recall volley at 50 Hz on both groups (Fig.~\ref{fig-da}a). The recall volley is identical for both cues and evokes presynaptic spikes that can read out a cue stored in synaptic states. The target output is $+1$ after cue A and $-1$ after cue B. Poisson background input (2 Hz on the cue groups and 20 Hz on four additional noise channels) is present throughout the trial. We use two initialisations of the input weights $W_{A}^{\mathrm{in}}$ and $W_{B}^{\mathrm{in}}$; all other weights are random in both. In the \emph{random initialisation}, they are independent ($W_{A}^{\mathrm{in}} \neq W_{B}^{\mathrm{in}}$), so that, by chance, some neurons already prefer one cue (Fig.~\ref{fig-da}b). In the \emph{matched initialisation}, the input weights of group B are copied from those of group A ($W_{A}^{\mathrm{in}} = W_{B}^{\mathrm{in}}$), so that no neuron prefers either cue and a cue preference must be established by learning.

\emph{Delayed XOR.} To test STPx e-prop without a constructed initialisation, we use two simultaneous 50 ms cues at 80 Hz, each at input group A or B (eight channels per group), with cue 1 on channels 1--4 and cue 2 on channels 5--8 of its group, followed by the delay, a recall volley on all cue channels, and the background input of delayed association (Fig.~\ref{fig-xor}a). The target output is $+1$ when the two cues are on different groups and $-1$ when they are on the same group. Neither cue alone predicts the target, so random input weights give no pre-existing advantage, and the memory must be written during learning.

\emph{Evidence accumulation.} We also use the evidence-accumulation task of~\cite{Bellec2020-yh}, in which seven cues of 100 ms are followed by a 1050 ms delay and the network reports which side received more cues, with facilitating recurrent synapses ($U = 0.025$, $\tau_{F} = 2$ s).

\emph{Training.} By default, networks of $n = 100$ neurons are trained with Adam~\cite{kingma2015-adam} (learning rate $3 \cdot 10^{-3}$, batch size 32; 10000 updates on delayed XOR and 3000 on the other tasks) and evaluated on held-out trials. Each condition is run with 10 seeds, and paired comparisons use a two-sided Wilcoxon signed-rank test across seeds; when all 10 seeds favour the same condition, $p = 0.002$, the smallest attainable value.

\subsection{Eligibility propagation (e-prop)}

Weights are learned by gradient descent on the loss, $\Delta W_{j i} = -\eta \, \dd E / \dd W_{j i}$. E-prop~\cite{Bellec2020-yh} approximates this gradient online: it is factorised into a synapse-local eligibility trace, computed forward in time, and a learning signal, approximated from the instantaneous, broadcast output error. As in~\cite{Bellec2020-yh}, all gradients, including those of BPTT, are taken with the surrogate derivative $\psi_{j}^{t}$ of $z_{j}^{t}$ with respect to $v_{j}^{t}$~\cite{neftci2019-surrogate}, and with the reset term of Eq.~\eqref{eq-lif} omitted from the Jacobian; ``exact'' below means exact for this gradient.

The weight $W_{j i}$, from presynaptic neuron $i$ to postsynaptic neuron $j$, affects the loss only through the spikes of neuron $j$, so the gradient factorises exactly into
\begin{equation}
  \frac{\dd E}{\dd W_{j i}} = \sum_{t} \frac{\dd E}{\dd z_{j}^{t}} \, \co{e_{j i}^{t}}, \quad \co{e_{j i}^{t}} = \psi_{j}^{t} \epsilon_{j i}^{t},
\label{eq-factor}
\end{equation}
where $e_{j i}^{t} = \dd z_{j}^{t} / \dd W_{j i}$ is the eligibility trace and $\epsilon_{j i}^{t} = \dd v_{j}^{t} / \dd W_{j i}$. The total derivative $\dd E / \dd z_{j}^{t}$ depends on future network activity, so e-prop replaces it by the partial derivative, the direct effect of the spike through the readout:
\begin{equation}
  \frac{\partial E}{\partial z_{j}^{t}} = \sum_{t' \geq t} \frac{\partial E}{\partial y^{t'}} \frac{\partial y^{t'}}{\partial z_{j}^{t}} = W_{j}^{\mathrm{out}} \sum_{t' \geq t} \cl{\delta^{t'}} (1 - \kappa) \kappa^{t' - t} ,
\label{eq-direct}
\end{equation}
which discards every effect of the spike through other neurons. The sum over $t'$ arises because the readout low-pass filters the spikes (Eq.~\eqref{eq-readout}), so that a spike affects all later outputs. Inserting Eq.~\eqref{eq-direct} into Eq.~\eqref{eq-factor} and exchanging the order of the two sums applies this filter to the eligibility trace instead~\cite{Bellec2020-yh}: the error $\delta^{t}$ then multiplies the low-pass-filtered trace $\bar{e}_{j i}^{t} = (1 - \kappa) \sum_{t' \leq t} \kappa^{t - t'} e_{j i}^{t'}$. Each synapse thus runs two forward recursions,
\begin{equation}
\begin{aligned}
  \co{\epsilon_{j i}^{t}} &= \alpha \epsilon_{j i}^{t-1} + (1 - \alpha) q_{j i}^{t}, \\
  \co{\bar{e}_{j i}^{t}} &= \kappa \bar{e}_{j i}^{t-1} + (1 - \kappa) \psi_{j}^{t} \epsilon_{j i}^{t},
\end{aligned}
\label{eq-traces}
\end{equation}
and the online update is
\begin{equation}
  \Delta W_{j i} = -\eta \sum_{t} \cl{\delta^{t}} \, W_{j}^{\mathrm{out}} \, \co{\bar{e}_{j i}^{t}} .
\label{eq-eprop}
\end{equation}
The trace thus records how much the weight contributed to the recent spiking of neuron $j$, and the learning signal whether more spiking of $j$ would reduce the error. With static synapses, $q_{j i}^{t} = z_{i}^{t-1}$ and $\epsilon_{j i}$ is the trace of~\cite{Bellec2020-yh}. We write $\bar{e}_{j i} = \mathcal{F}_{j} [q_{j i}]$: the filter $\mathcal{F}_{j} [\,\cdot\,]$ of Eq.~\eqref{eq-traces} passes its argument through the membrane filter, gates it with $\psi_{j}$, and low-pass filters it as the readout does.

\emph{Naming.} Throughout, e-prop denotes Eq.~\eqref{eq-eprop} applied to the network with STP, so that the release $q_{j i}$ enters its eligibility trace. With the readout weight $W_{j}^{\mathrm{out}}$ as feedback it is \emph{symmetric e-prop}; with a fixed random weight $B_{j}$ in its place (random e-prop~\cite{Bellec2020-yh}), it requires no weight transport, and we call it \emph{e-prop}. Neither assigns credit through the states of the outgoing synapses. As a control, we also train e-prop with the trace of a static synapse, that is, with the presynaptic spike $z_{i}$ in place of $q_{j i}$ in the eligibility trace of the recurrent synapses; the network itself keeps its STP.

\section{STP-extended e-prop}\label{sec-stpx}

In this section, we derive STPx e-prop, which adds to the e-prop update a second eligibility trace with its own feedback:
\begin{equation}
  \Delta W_{j i} = -\eta \sum_{t} \cl{\delta^{t}} (B_{j} \, \co{\bar{e}_{j i}^{t}} + \gamma \rho_{0} \operatorname{sgn}(B_{j}) \, \ct{\bar{r}_{j i}^{t}}) .
\label{eq-local}
\end{equation}
The first term is e-prop with the random feedback weight $B_{j}$. In the second, $\bar{r}_{j i}$ is a filtered trace of how the release of neuron $j$ onto its targets depends on $W_{j i}$ through the state of its outgoing synapses, and its feedback is the sign of the same random weight, scaled by a fixed gain $\gamma \rho_{0}$. From the exact gradient we first derive its form with exact feedback, STPx e-prop + $b_{j} \delta$ (Eq.~\eqref{eq-rule}), in which $W_{j}^{\mathrm{out}}$ and a weight $b_{j}$ take the places of $B_{j}$ and $\gamma \rho_{0} \operatorname{sgn}(B_{j})$, and then replace both with these local signals (Section~\ref{sec-local}).

\subsection{Credit paths and gradient factorisation}

A spike $z_{j}^{t}$ reaches the loss along three paths (Fig.~\ref{fig-model}b): (a) directly, through the readout; (b1) through the release $q_{k j}^{t+1}$ that it evokes onto each target $k$ at the next step; and (b2) through the state $\boldsymbol{s}_{j} = (u_{j}, x_{j})$ that it leaves in the outgoing synapses of $j$, which modulates later releases and is where a synaptic memory is written. E-prop retains only (a): (b1) and (b2) pass through other neurons and require their credit $\lambda_{k}^{t} = \dd E / \dd v_{k}^{t}$, which BPTT computes backward in time.

To recover these paths, we extend Eq.~\eqref{eq-factor} from the spikes of neuron $j$ to a group of variables that $W_{j i}$ influences without passing through another neuron: its spikes, its membrane potential, and $\boldsymbol{s}_{j}$. Every path from $W_{j i}$ to the loss first leaves this group either at a spike into the readout, or at a release into the membrane potential of a target, so that, without approximation,
\begin{equation}
  \frac{\dd E}{\dd W_{j i}} = \sum_{t} \Biggl[ \underbrace{\cl{\frac{\partial E}{\partial z_{j}^{t}}} \co{e_{j i}^{t}}}_{\text{(a)}} + \sum_{k} \underbrace{(1 - \alpha) W_{k j} \lambda_{k}^{t} \frac{\dd q_{k j}^{t}}{\dd W_{j i}}}_{\text{(b1) + (b2)}} \Biggr] .
\label{eq-exact}
\end{equation}
As in Eq.~\eqref{eq-factor}, each term is a credit times a forward sensitivity, and the only backward quantity is the credit $\lambda_{k}^{t} = \dd E / \dd v_{k}^{t}$.

\subsection{Terminal-state eligibility trace}

Because $W_{j i}$ reaches every variable of the group without leaving it, both sensitivities in Eq.~\eqref{eq-exact} can be propagated forward in time. For the spikes this is $e_{j i}^{t}$ of Eq.~\eqref{eq-factor}. The release onto any target $k$ depends on $W_{j i}$ through the spike that evokes it and through the synaptic state, so that
\begin{equation}
  \frac{\dd q_{k j}^{t}}{\dd W_{j i}} = \underbrace{q_{z}^{t} \co{e_{j i}^{t-1}}}_{\text{(b1)}} + \underbrace{\boldsymbol{q}_{s}^{t} \cdot \ct{\boldsymbol{S}_{j i}^{t}}}_{\ct{r_{j i}^{t}} \text{(b2)}} ,
\label{eq-split}
\end{equation}
where $\boldsymbol{S}_{j i}^{t} = \dd \boldsymbol{s}_{j}^{t} / \dd W_{j i}$ is the terminal-state trace. Differentiating one TM step (Eq.~\eqref{eq-tm}) gives a linear forward recursion for it, driven by $e_{j i}$ (Eq.~\eqref{eq-vars}). Its coefficients are the partial derivatives of Eq.~\eqref{eq-tm} at $(u, x) = \boldsymbol{s}_{j}^{t}$ and $z = z_{j}^{t-1}$, with $u^{+} = u + U (1 - u) z$:
\begin{equation}
\begin{aligned}
  J_{s}^{t} & = \frac{\partial \boldsymbol{s}_{j}^{t+1}}{\partial \boldsymbol{s}_{j}^{t}} = \begin{pmatrix} d_{F} (1 - U z) & 0 \\
  -d_{D} x z (1 - U z) & d_{D} (1 - u^{+} z) \end{pmatrix}, \\
  \boldsymbol{j}_{z}^{t} & = \frac{\partial \boldsymbol{s}_{j}^{t+1}}{\partial z_{j}^{t-1}} = \begin{pmatrix} d_{F} U (1 - u) \\
  -d_{D} x (u^{+} + z U (1 - u)) \end{pmatrix}, \\
  \boldsymbol{q}_{s}^{t} & = \partial q_{k j}^{t} / \partial \boldsymbol{s}_{j}^{t} = (A_{0} x z (1 - U z), \, A_{0} u^{+} z), \\
  q_{z}^{t} & = \partial q_{k j}^{t} / \partial z_{j}^{t-1} = A_{0} x (u^{+} + z U (1 - u)) .
\end{aligned}
\label{eq-jac}
\end{equation}
All coefficients depend only on neuron $j$ and its outgoing synapses, and are shared by all its incoming synapses.

Unlike the adaptation component of the eligibility vector in~\cite{Bellec2020-yh}, $\boldsymbol{S}_{j i}$ does not enter $e_{j i}$: the spike of neuron $j$ does not depend on the state of its own outgoing synapses.

\subsection{Online approximation}\label{sec-steps}

Term (a) of Eq.~\eqref{eq-exact} is the e-prop update of Eq.~\eqref{eq-eprop}. STP-extended e-prop (STPx e-prop) adds the synaptic memory term obtained by inserting (b2) of Eq.~\eqref{eq-split} into Eq.~\eqref{eq-exact}, $\sum_{t, k} (1 - \alpha) W_{k j} \lambda_{k}^{t} r_{j i}^{t}$, and removes its backward quantity $\lambda_{k}^{t}$ in three steps. The first two are approximations already made by e-prop~\cite{Bellec2020-yh}; only the third is new.

\emph{Step 1: omit the immediate release.} We drop (b1), as e-prop does. It acts through the release at the step after the spike and not through the stored state, so it carries no credit for writing a memory held in the synapse.

\emph{Step 2: e-prop at the target.} With the reset omitted, $\lambda_{k}^{t} = \psi_{k}^{t} \, \dd E / \dd z_{k}^{t} + \alpha \lambda_{k}^{t+1} = \sum_{t' \geq t} \alpha^{t' - t} \psi_{k}^{t'} \, \dd E / \dd z_{k}^{t'}$. Inserting this and exchanging the order of summation moves the membrane filter from $\lambda_{k}$ onto $r_{j i}$, which is exact:
\begin{equation}
  \sum_{t} (1 - \alpha) \lambda_{k}^{t} \ct{r_{j i}^{t}} = \sum_{t} \frac{\dd E}{\dd z_{k}^{t}} \, \psi_{k}^{t} \ct{\mu_{j i}^{t}},
\label{eq-unroll}
\end{equation}
where $\mu_{j i}$ is $r_{j i}$ passed through the membrane filter (Eq.~\eqref{eq-vars}). Applying the e-prop approximation at the target, $\dd E / \dd z_{k}^{t} \approx \partial E / \partial z_{k}^{t}$, and moving the readout filter onto the trace as in Eq.~\eqref{eq-traces} yields the one-hop credit
\begin{equation}
  \text{(b2)} \approx \sum_{t} \cl{\delta^{t}} \sum_{k} W_{k j} W_{k}^{\mathrm{out}} \, \ct{\mathcal{F}_{k} [r_{j i}]^{t}},
\label{eq-hop}
\end{equation}
where $\mathcal{F}_{k}$ is the filter of Eq.~\eqref{eq-traces} with the gate of the target, $\psi_{k}$, in place of $\psi_{j}$.
Each target is thus credited as e-prop credits it, by its readout weight and the current error; like e-prop, Eq.~\eqref{eq-hop} covers a single hop and omits paths through further neurons. It is exact when the only recurrent connection is $j \to k$, since $z_{k}$ then reaches the loss only through the readout.

\emph{Step 3: gate swap.} Eq.~\eqref{eq-hop} is not local: the gate $\psi_{k}$ of the target, the only part of $\mathcal{F}_{k}$ specific to $k$, sits between the two filters, so each synapse would require one filtered trace per target. Evaluating the filter with the gate of the sender, replacing $\mathcal{F}_{k}$ with $\mathcal{F}_{j}$, makes it identical for all targets, and the sum over targets reduces to a single number per neuron:
\begin{equation}
\begin{aligned}
  \text{(b2)} & \approx \sum_{t} \cl{\delta^{t}} \, b_{j} \, \ct{\mathcal{F}_{j} [r_{j i}]^{t}}, \quad b_{j} = \sum_{k} W_{k j} W_{k}^{\mathrm{out}} .
\end{aligned}
\label{eq-swap}
\end{equation}
The swap assumes that the targets are sensitive when the sender is; $b_{j}$ then summarises them in one number, the strength and sign with which the releases of $j$ are connected to the output. Because $\psi \geq 0$, the swap preserves the sign of every contribution $\psi^{t} \mu_{j i}^{t}$ to the filter and only reweights them in time; it is a weaker substitute when $j$ and its targets are sensitive at different times. On delayed XOR with facilitating synapses, the swap reduces the cosine similarity between the STP term and the gradient of BPTT through the synaptic state from 0.67--1.00 to 0.24--0.71 over training.

\subsection{The rule with exact feedback}\label{sec-rule}

Combining Eqs.~\eqref{eq-eprop} and \eqref{eq-swap}, the weight is updated by the output error $\cl{\delta^{t}}$ times two eligibility traces, each with its own feedback weight:
\begin{equation}
\begin{aligned}
  \Delta W_{j i} & = -\eta \sum_{t} \cl{\delta^{t}} (W_{j}^{\mathrm{out}} \co{\bar{e}_{j i}^{t}} + b_{j} \ct{\bar{r}_{j i}^{t}}), \\
  \co{\bar{e}_{j i}} & = \mathcal{F}_{j} [q_{j i}], \quad \ct{\bar{r}_{j i}} = \mathcal{F}_{j} [r_{j i}] .
\end{aligned}
\label{eq-rule}
\end{equation}
The first term is symmetric e-prop and credits $W_{j i}$ for the spikes it causes in $j$; the second credits it for the synaptic state that those spikes leave behind. Here, $b_{j}$ sums the readout weights of the targets of $j$, weighted by the synapses onto them; equivalently, $b_{j} \delta^{t} = \sum_{k} W_{k j} L_{k}^{t}$ sums their e-prop learning signals $L_{k}^{t} = W_{k}^{\mathrm{out}} \delta^{t}$. For an input weight $W_{j a}^{\mathrm{in}}$, the input spike $z_{a}^{\mathrm{in}, t}$ replaces $q_{j i}^{t}$ in Eq.~\eqref{eq-traces}, and therefore in $\epsilon$, $e$ and $\boldsymbol{S}$.

Each synapse $i \to j$ maintains five variables, updated forward in time from quantities available at neuron $j$: $\co{\bar{e}_{j i}}$ of e-prop (Eq.~\eqref{eq-traces}; $\co{\epsilon_{j i}}$ is shared by all targets of $i$) and four more, the two components of the terminal-state trace $\ct{\boldsymbol{S}_{j i}}$, read out as $\ct{r_{j i}}$, and $\ct{\mu_{j i}}$ and $\ct{\bar{r}_{j i}}$, which filter $r_{j i}$ as $\epsilon_{j i}$ and $\bar{e}_{j i}$ filter the release $q_{j i}$:
\begin{equation}
\begin{aligned}
  \ct{\boldsymbol{S}_{j i}^{t+1}} &= J_{s}^{t} \boldsymbol{S}_{j i}^{t} + \boldsymbol{j}_{z}^{t} \psi_{j}^{t-1} \epsilon_{j i}^{t-1}, \quad \ct{r_{j i}^{t}} = \boldsymbol{q}_{s}^{t} \cdot \boldsymbol{S}_{j i}^{t}, \\
  \ct{\mu_{j i}^{t}} &= \alpha \mu_{j i}^{t-1} + (1 - \alpha) r_{j i}^{t}, \\
  \ct{\bar{r}_{j i}^{t}} &= \kappa \bar{r}_{j i}^{t-1} + (1 - \kappa) \psi_{j}^{t} \mu_{j i}^{t} .
\end{aligned}
\label{eq-vars}
\end{equation}
with $\boldsymbol{S}_{j i}^{0} = \boldsymbol{0}$. Per time step, the STP term costs each synapse one $2 \times 2$ matrix--vector product and two scalar filters. If only a fraction $p$ of the neurons have STP, only their incoming synapses carry the four additional variables, and a synapse stores $1 + 4 p$ variables on average.

\subsection{Local feedback: STPx e-prop}\label{sec-local}

We call Eq.~\eqref{eq-rule}, with the symmetric feedback $W_{j}^{\mathrm{out}}$ and the exact $b_{j}$, STPx e-prop + $b_{j} \delta$. Its feedback is not local: $b_{j}$ requires the weights from $j$ to its targets and their readout weights (dashed orange arrow in Fig.~\ref{fig-model}b), a weight-transport problem~\cite{lillicrap2016-random}. We replace it in three steps: the readout weight of the e-prop term by a random weight, the sign of $b_{j}$ by the sign of that random weight, and its scale by a fixed gain. Each step is tested on delayed association in Section~\ref{sec-needs}.

\emph{Only the sign has to be right.} The STP term of neuron $j$ in Eq.~\eqref{eq-rule} can be written as $b_{j} \, \boldsymbol{a}_{j}$, where $\boldsymbol{a}_{j}$ has one component $\sum_{t} \delta^{t} \, \bar{r}_{j i}^{t}$ per incoming synapse $i$ of $j$. The terms of different neurons occupy disjoint coordinates, so a substitute $\tilde{b}_{j}$ for $b_{j}$ turns the STP term of all recurrent weights, $\boldsymbol{G} = (b_{j} \boldsymbol{a}_{j})_{j}$, into a $\tilde{\boldsymbol{G}} = (\tilde{b}_{j} \boldsymbol{a}_{j})_{j}$ whose inner product with it is
\begin{equation}
  \langle \tilde{\boldsymbol{G}}, \boldsymbol{G} \rangle = \sum_{j} \tilde{b}_{j} \, b_{j} \, \lVert \boldsymbol{a}_{j} \rVert^{2} .
\label{eq-inner}
\end{equation}
As long as this inner product is positive, an update along $\tilde{\boldsymbol{G}}$ also moves the weights along $\boldsymbol{G}$~\cite{lillicrap2016-random}. For sign-only feedback, $\tilde{b}_{j} = \rho \, s_{j}$ with $s_{j} = \pm 1$ and the scale $\rho = \operatorname{rms}_{j} b_{j}$, it is positive if and only if the fraction of correct signs, weighted by $|b_{j}| \, \lVert \boldsymbol{a}_{j} \rVert^{2}$, exceeds one half.

\emph{Random feedback and its sign.} As in e-prop, the random weight $B_{j}$ replaces $W_{j}^{\mathrm{out}}$ in the first term. The matching feedback for the STP term, $\sum_{k} W_{k j} B_{k}$, would still require one-hop communication. We instead use the sign of the neuron's own random weight, $s_{j} = \operatorname{sgn}(B_{j})$, so that the STP term requires neither the readout weights nor the forward weights. Learning through $B$ can align the vector of the $b_{j}$ with $B$, as in feedback alignment~\cite{lillicrap2016-random}; we measure this alignment in Section~\ref{sec-needs}.

\begin{figure}[t]
\centering
\begin{tabular}{@{}c@{\hspace{10pt}}c@{}}
\includegraphics[width=154pt]{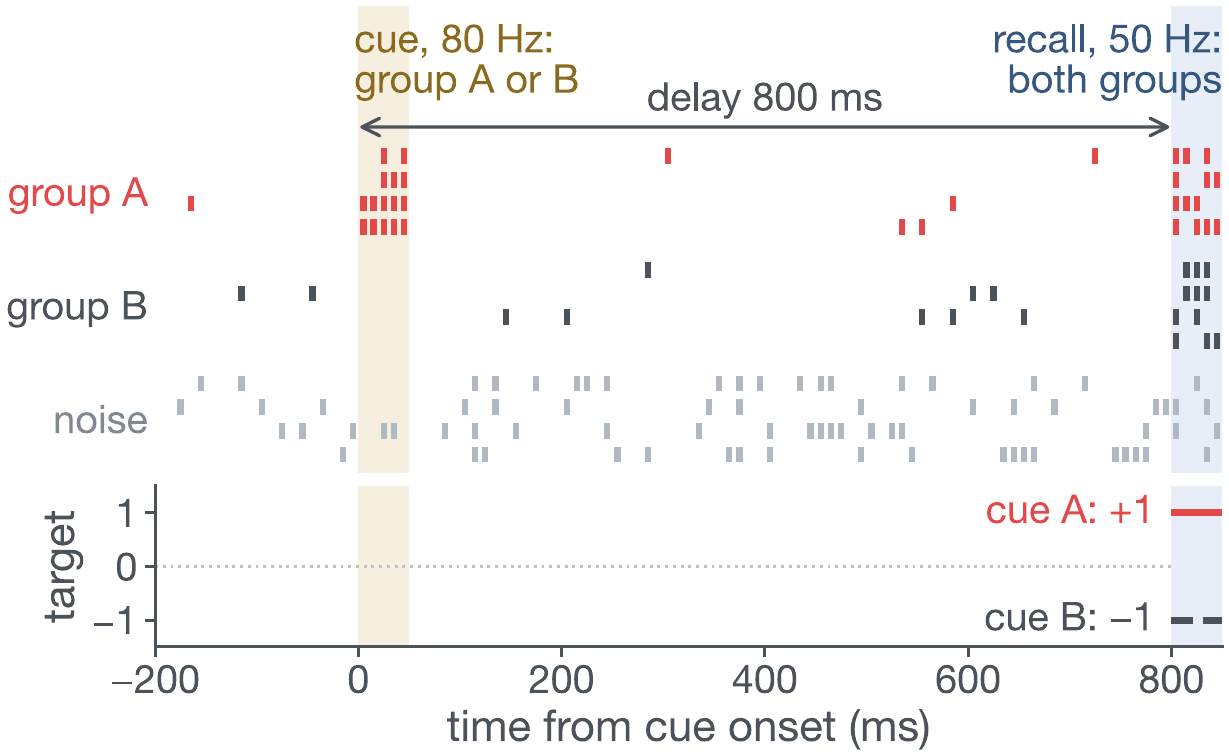} & \includegraphics[width=87pt]{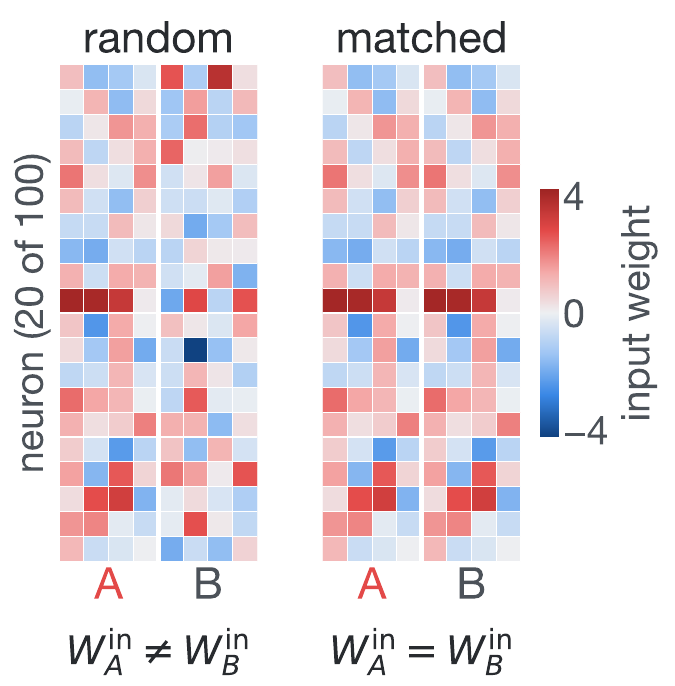}\\[-1pt]
{\footnotesize (a) one cue-A trial} & {\footnotesize (b) initial input weights}
\end{tabular}\\[4pt]
\includegraphics[width=\columnwidth]{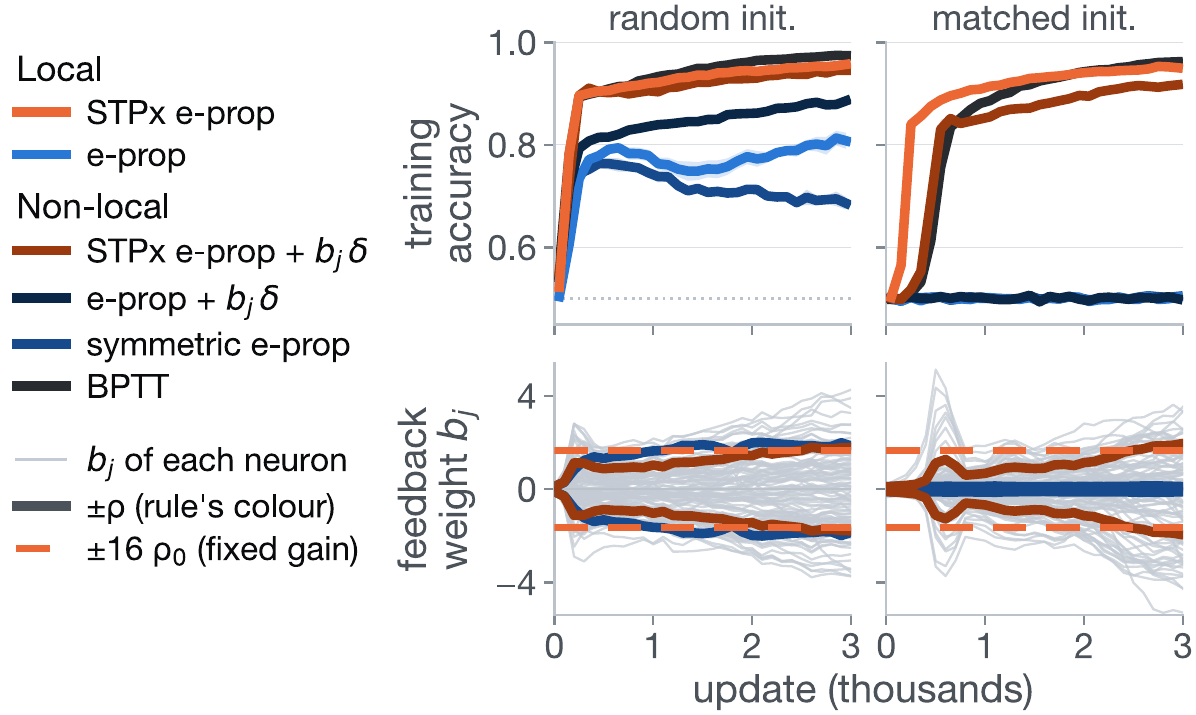}\\[-1pt]
{\footnotesize (c) training accuracy and feedback weights}
\caption{Delayed association. (a) Input spikes and target output in one cue-A trial. (b) Initial input weights from cue groups A and B in one network, with the random and the matched initialisation. (c) Top: training accuracy from the random (left) and the matched initialisation (right; mean $\pm$ s.e. over 10 seeds). Apart from BPTT, the non-local rules use the readout weights as feedback. Bottom: feedback weights $b_{j}$ of one network trained with STPx e-prop + $b_{j} \delta$ (thin), their root mean square $\pm \rho$ (thick; blue: symmetric e-prop from the same initial weights), and the fixed gain $\pm 16 \rho_{0}$ of STPx e-prop (dashed).}
\label{fig-da}
\end{figure}

\emph{Fixed gain.} The scale $\rho$ is a single number for the whole network, which a global signal such as a neuromodulator could broadcast; the sign, in contrast, must be specific to each neuron. Computing $\rho$ still requires the weights, so we replace it with a constant $\gamma \rho_{0}$, where $\rho_{0}$ is $\rho$ before training and $\gamma = 16$ by default (dashed in Fig.~\ref{fig-da}c, bottom).

\emph{STPx e-prop.} The three replacements give the fully local rule of Eq.~\eqref{eq-local}: besides the traces of Eqs.~\eqref{eq-traces} and \eqref{eq-vars}, a synapse requires only the random weight $B_{j}$ of its neuron and the broadcast error $\delta^{t}$.

\section{Results}

We introduced STPx e-prop (Section~\ref{sec-stpx}), which extends e-prop with a second eligibility trace, of the outgoing synaptic state, and thereby assigns credit for writing memories into short-term synaptic states. Here, we evaluate it on three tasks (Section~\ref{sec-tasks}).

\subsection{STPx e-prop writes a synaptic memory}\label{sec-write}

\begin{figure}[t]
\centering
\includegraphics[width=\columnwidth]{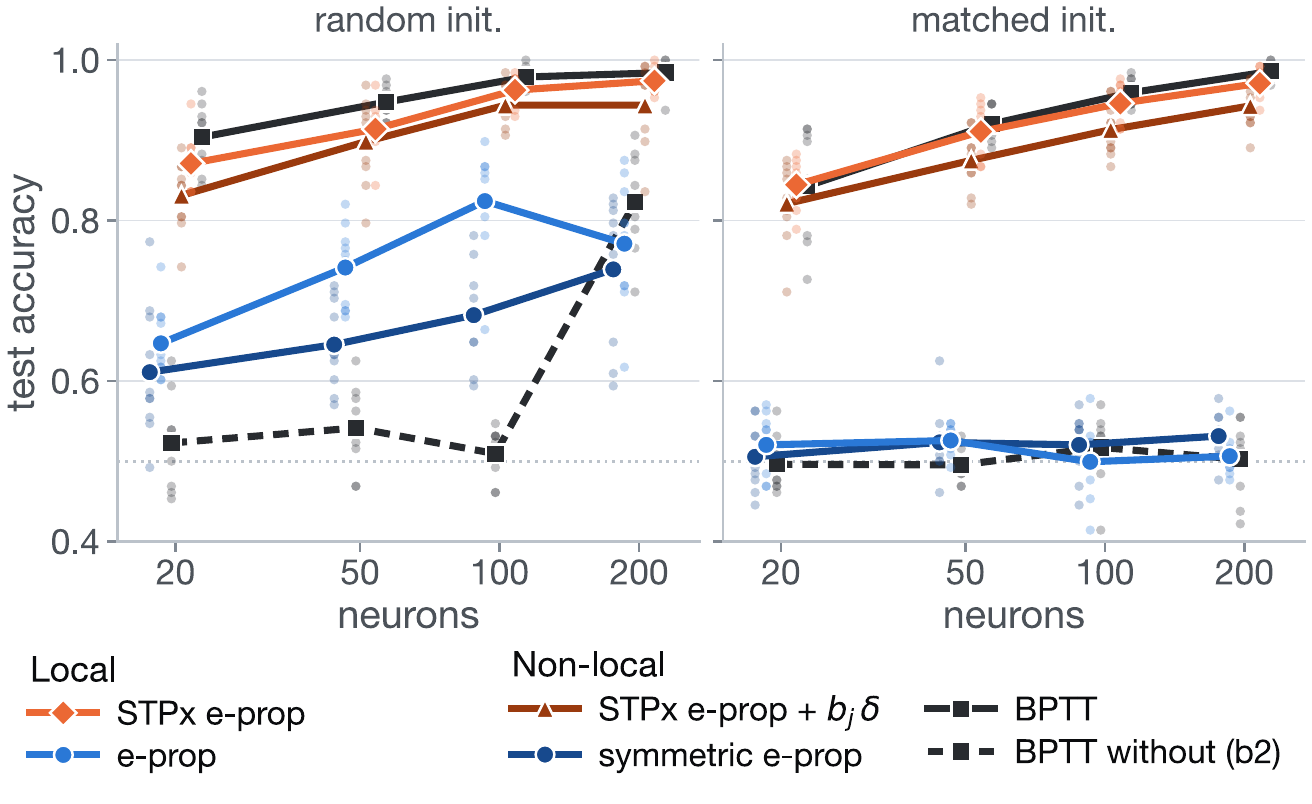}
\caption{Delayed association as a function of network size. Test accuracy from the random (left) and the matched initialisation (right), with facilitating recurrent synapses (dots: 10 seeds; lines: mean). BPTT without (b2): BPTT with the path through the synaptic state removed from the gradient.}
\label{fig-size}
\end{figure}

For the delayed association task (Fig.~\ref{fig-da}a), the matched initialisation (Fig.~\ref{fig-da}b) tests whether a rule writes the memory, that is, whether it changes the input and recurrent weights so that each cue leaves its own state in the synapses through the delay: no neuron prefers either cue before learning. From this initialisation, STPx e-prop learns delayed association with a test accuracy of 0.95, not significantly below BPTT (0.96, $p = 0.16$), whereas e-prop remains at chance (0.50; Fig.~\ref{fig-da}c, Table~\ref{tab-local}). From the random initialisation, STPx e-prop reaches 0.96 (BPTT 0.98, $p = 0.078$), and e-prop reaches 0.82 by exploiting the cue preference already present in the input weights. STPx e-prop is also not below STPx e-prop + $b_{j} \delta$ from either initialisation (0.94 and 0.91; differences of 0.02 and 0.03 in favour of STPx e-prop, $p = 0.053$ and $0.076$). The same holds at every network size (Fig.~\ref{fig-size}). From the matched initialisation, STPx e-prop is within 0.02 of BPTT from 20 to 200 neurons (0.84--0.97, compared with 0.84--0.99; $p \geq 0.062$), and e-prop remains at chance (0.50--0.53).

Symmetric feedback does not change this. From the matched initialisation, symmetric e-prop remains at chance at every network size (0.51--0.53), whereas STPx e-prop + $b_{j} \delta$ remains within 0.02--0.05 of BPTT (Fig.~\ref{fig-size}). From the random initialisation, both e-prop variants gain only from the pre-existing cue preference (0.61--0.74 for symmetric e-prop and 0.65--0.82 for e-prop), and symmetric e-prop loses part of it during training (training accuracy 0.76 after 500 updates and 0.69 at the end; Fig.~\ref{fig-da}c). This learning depends on the release in the eligibility trace: with the presynaptic spike in its place, e-prop remains at chance from the random initialisation at 50, 100 and 200 neurons (0.51--0.53, compared with 0.74--0.82; 10 of 10 seeds lower at each size, $p = 0.002$). From the matched initialisation both remain at chance (0.50--0.53), so the release in the trace lets e-prop read a stored cue, but it does not learn to write one here. Path (b2) is necessary for writing the memory: with only (b2) removed, BPTT remains at chance from the matched initialisation at every network size (0.50--0.52, compared with 0.84--0.99; $p = 0.002$ at each size; Fig.~\ref{fig-size}).

\begin{figure*}[t]
\centering
\begin{tabular}{@{}c@{\hspace{14pt}}c@{}}
\raisebox{-0.5\height}{\includegraphics[width=169pt]{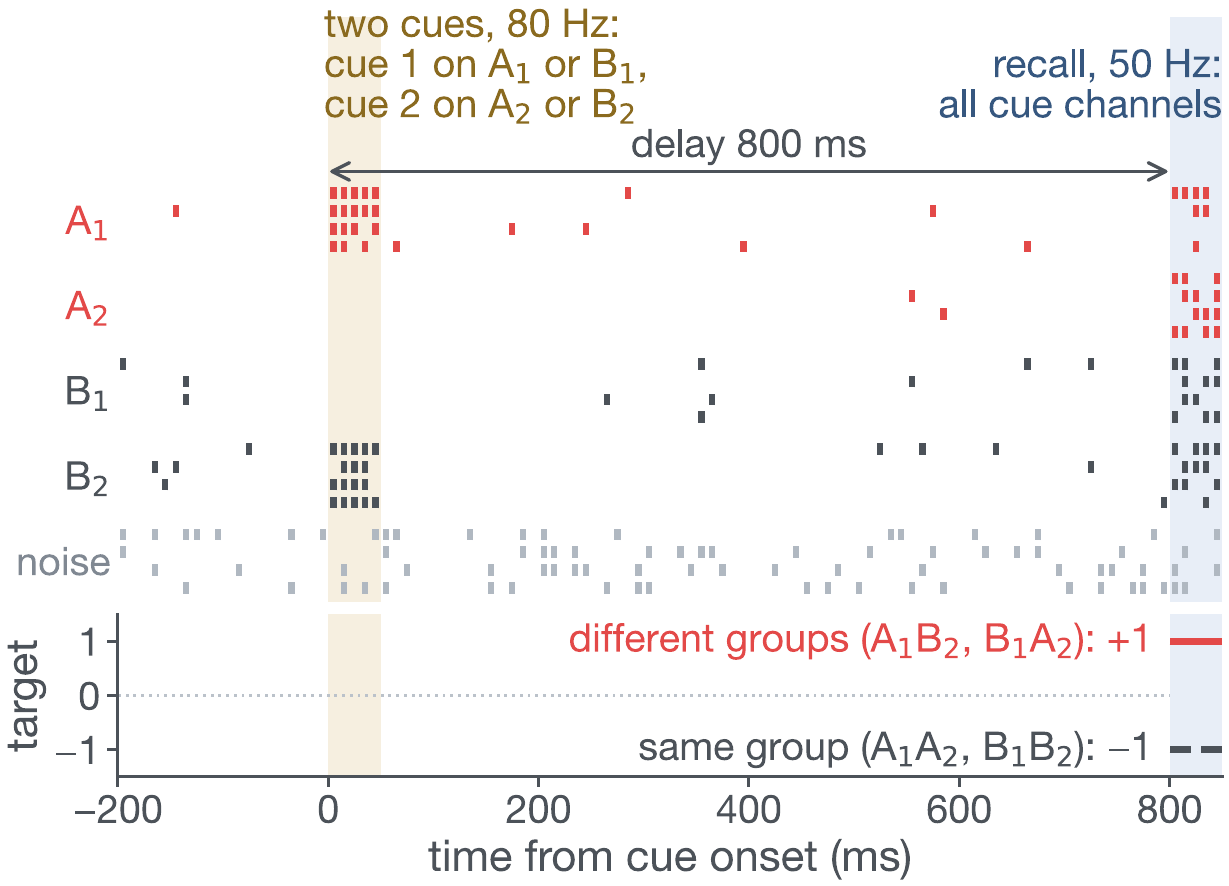}} &
\raisebox{-0.5\height}{\includegraphics[width=236pt]{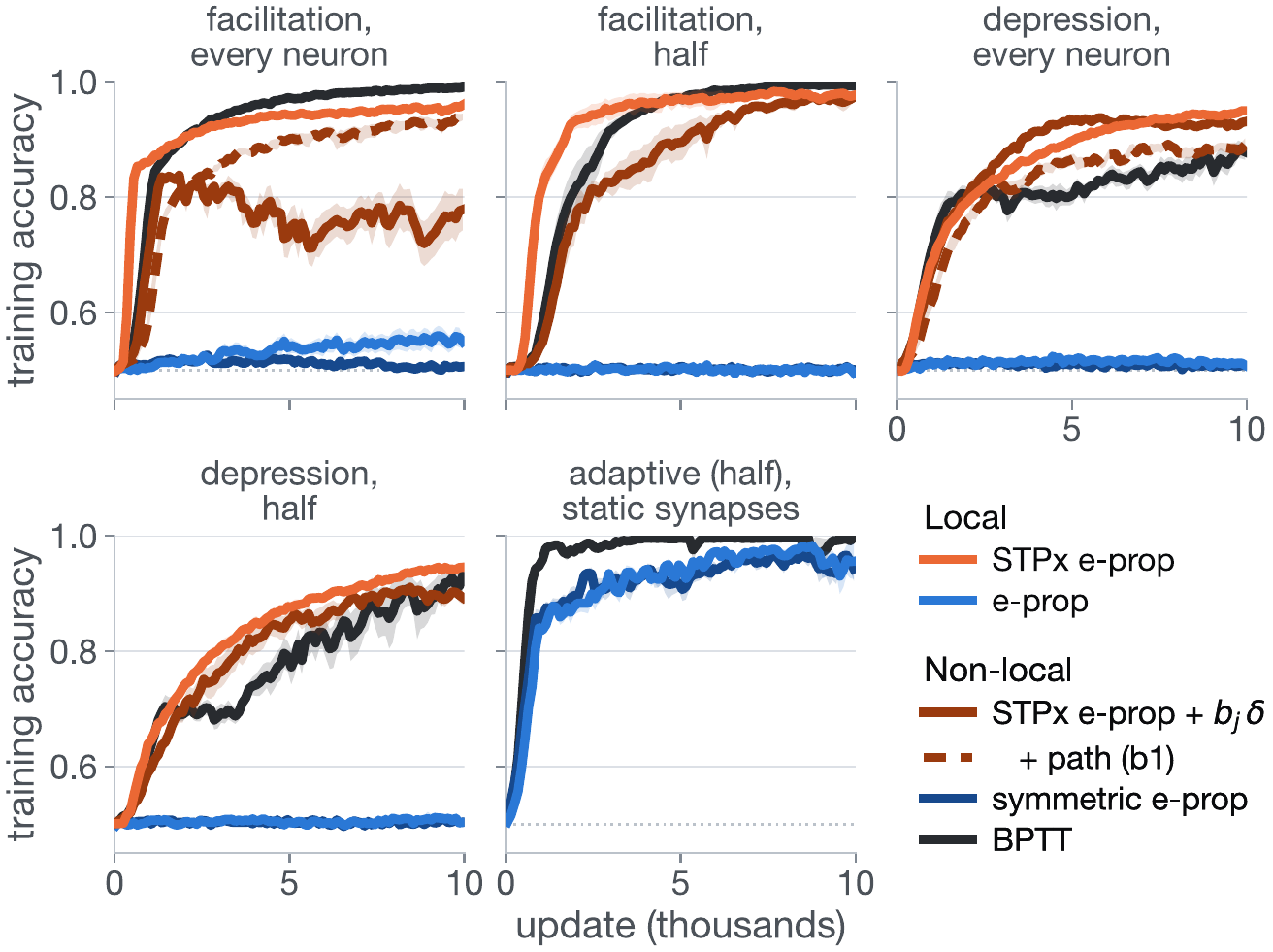}}\\[10pt]
{\footnotesize (a) one trial of delayed XOR} & {\footnotesize (b) learning curves}
\end{tabular}
\caption{Delayed XOR. (a) Input spikes and target output in one cue-different trial. The target is $+1$ when the cues are on different groups (A\textsubscript{1}B\textsubscript{2} or B\textsubscript{1}A\textsubscript{2}) and $-1$ when they are on the same group. (b) Training accuracy (100 neurons, random initialisation; mean $\pm$ s.e. over 10 seeds). Apart from BPTT, the non-local rules use the readout weights as feedback. Facilitation or depression, in every neuron or in half; dashed, STPx e-prop + $b_{j} \delta$ with path (b1) added; last panel, static synapses with half the neurons adaptive.}
\label{fig-xor}
\end{figure*}

\subsection{What the rule requires: trace, sign and gain}\label{sec-needs}

\emph{Trace versus feedback.} To separate the contribution of the terminal-state trace from that of the additional feedback, we trained symmetric e-prop with the feedback $b_{j} \delta$ added to its learning signal but without the STP trace (e-prop + $b_{j} \delta$). This control exploits a pre-existing cue preference (0.89, compared with 0.68 for symmetric e-prop, from the random initialisation) but does not learn the task from the matched initialisation (0.51). The feedback thus improves the readout of a pre-existing cue, whereas writing a memory requires the trace.

\begin{table}[b]
\caption{Test accuracy on delayed association as the non-local quantities of STPx e-prop + $b_{j} \delta$ are replaced (100 neurons, mean over 10 seeds). Upper row: e-prop term with $W_{j}^{\mathrm{out}}$; lower rows: with $B_{j} \delta$. Blue: e-prop; orange: STPx e-prop, which requires no weight transport.}
\label{tab-local}
\centering
\footnotesize
\begin{tabular*}{\columnwidth}{@{}l@{\extracolsep{\fill}}cc@{}}
\toprule
 & rand. & match. \\
\midrule
\ctd{STPx e-prop + $b_j \delta$} & 0.94 & 0.91 \\
\midrule
\co{e-prop, $B_j \delta$} & 0.82 & 0.50 \\
\co{e-prop, $B_j \delta$, spike in the trace} & 0.53 & 0.50 \\
\ctd{$B_j \delta$ + exact $b_j$} & 0.96 & 0.92 \\
\ctd{$B_j \delta$ + $\sum_k W_{kj} B_k$} & 0.97 & 0.95 \\
\ct{$B_j \delta$ + $\operatorname{sgn}(B_j) \times 16 \rho_0$} & 0.96 & 0.95 \\
\midrule
BPTT & 0.98 & 0.96 \\
\bottomrule
\end{tabular*}
\end{table}

\emph{Sign.} We replaced the non-local quantities of STPx e-prop + $b_{j} \delta$ one at a time (Section~\ref{sec-local}; Table~\ref{tab-local}); no replacement reduces accuracy significantly. With random feedback in the e-prop term, the sign of the neuron's own random weight performs as well as the exact $b_{j}$ (0.96 and 0.95, compared with 0.96 and 0.92; $p = 0.36$ and $0.09$) and as the matching one-hop feedback $\sum_{k} W_{k j} B_{k}$ (0.97 and 0.95; $p \geq 0.56$). In the networks of Fig.~\ref{fig-da}c trained with STPx e-prop, the $|b_{j}|$-weighted fraction of neurons in which $b_{j}$ has the sign of $B_{j}$, a proxy for the weighted fraction of Eq.~\eqref{eq-inner}, rises from 53 \% before training to 92--98 \% from 300 updates on: learning through $B$ aligns the $b_{j}$ with it~\cite{lillicrap2016-random}.

\emph{Fixed gain.} During training with STPx e-prop + $b_{j} \delta$, $\rho$ grows about 20-fold from its initial value $\rho_{0}$, as the $b_{j}$ spread to both signs (Fig.~\ref{fig-da}c, bottom). With symmetric e-prop from the matched initialisation, which does not learn the task, $\rho$ stays near its initial value (1.2-fold). A constant of the order of the final $\rho$, $16 \rho_{0}$, performs as well as the exact $b_{j}$, which carries its own scale (Table~\ref{tab-local}). With gains of $\gamma = 1$, 4 and 64 in place of 16, STPx e-prop reaches 0.91, 0.94 and 0.85 from the random initialisation and 0.89, 0.92 and 0.84 from the matched one (0.96 and 0.95 at $\gamma = 16$).

\subsection{Delayed XOR}\label{sec-xor}

On delayed XOR (Fig.~\ref{fig-xor}a), symmetric e-prop and e-prop with facilitating recurrent synapses remain near chance (test accuracy 0.50 and 0.56), whereas STPx e-prop reaches 0.96 (10 of 10 seeds above both, $p = 0.002$), 0.03 below BPTT (0.99, $p = 0.002$; Fig.~\ref{fig-xor}b; Table~\ref{tab-xor-test}). Although its feedback is exact, STPx e-prop + $b_{j} \delta$ reaches only 0.76 and learns unevenly (0.20 below STPx e-prop, $p = 0.002$). The gate swap is not the cause: with the targets' own gates (Eq.~\eqref{eq-hop}), the rule does not improve (0.79, $p = 0.49$). Of the changes we tested, only one improves this rule: adding the immediate release, path (b1), which Step 1 of Section~\ref{sec-steps} omits, under the same approximations as the memory term raises the accuracy to 0.94 (10 of 10 seeds, $p = 0.002$; dashed in Fig.~\ref{fig-xor}b). This explains the deficit of STPx e-prop + $b_{j} \delta$ only in part, and with depression in every neuron the same term lowers accuracy (0.90, compared with 0.93; $p = 0.035$). Both STPx e-prop + $b_{j} \delta$ and the term (b1) require the exact $b_{j}$ and are therefore non-local.

At the same learning rate, STPx e-prop reaches a training accuracy of 0.80 in fewer updates than BPTT on delayed association with matched initialisation and on delayed XOR with facilitation (10 of 10 seeds each, $p = 0.002$; Figs.~\ref{fig-da}c and \ref{fig-xor}b), while BPTT reaches the higher final accuracy on delayed XOR.

With adaptive neurons and static synapses (Section~\ref{sec-network}), symmetric e-prop learns the task without an STP term (Fig.~\ref{fig-xor}b, last panel) and reaches 0.95, not significantly different from STPx e-prop on facilitating synapses (0.96, $p = 0.98$); we return to this in the Discussion. With facilitation in only half of the neurons, as for adaptation, STPx e-prop reaches 0.98, above symmetric e-prop with adaptive neurons ($p = 0.039$).

\begin{table}[b]
\caption{Test accuracy on delayed XOR (100 neurons, mean over 10 seeds). Facilitation, depression or adaptive neurons (static synapses) in all or half of the neurons.}
\label{tab-xor-test}
\centering
\footnotesize
\begin{tabular*}{\columnwidth}{@{}l@{\extracolsep{\fill}}ccccc@{}}
\toprule
 & \multicolumn{2}{c}{fac.} & \multicolumn{2}{c}{dep.} & adapt. \\
 & all & half & all & half & half \\
\midrule
\cod{symmetric e-prop} & 0.50 & 0.50 & 0.50 & 0.51 & 0.95 \\
\co{e-prop} & 0.56 & 0.51 & 0.51 & 0.51 & 0.94 \\
\ctd{STPx e-prop + $b_j \delta$} & 0.76 & 0.97 & 0.93 & 0.90 & -- \\
\ct{STPx e-prop} & 0.96 & 0.98 & 0.95 & 0.95 & -- \\
BPTT & 0.99 & 0.99 & 0.89 & 0.93 & 1.00 \\
\bottomrule
\end{tabular*}
\end{table}

\emph{STP variants.} The deficit of STPx e-prop + $b_{j} \delta$ is specific to facilitation in every neuron: with facilitation in half of the neurons and with depression in every neuron or in half, it reaches 0.97, 0.93 and 0.90, while symmetric e-prop remains at 0.50--0.51 (Table~\ref{tab-xor-test}). STPx e-prop does not differ from it with facilitation in half of the neurons (0.98, $p = 0.64$) and exceeds it with depression (0.95 in every neuron and in half; $p = 0.010$ and $0.002$). With depression in every neuron, both STPx rules exceed BPTT (0.93 and 0.95, compared with 0.89; $p = 0.002$), which is therefore not an upper bound here.

\subsection{Evidence accumulation}\label{sec-ea}

Evidence accumulation (Section~\ref{sec-tasks}) tests whether the rule works beyond the tasks designed for it. In networks with facilitating synapses and no adaptive neurons, STPx e-prop reaches 0.89, 0.92 and 0.94 at 50, 100 and 200 neurons and outperforms e-prop (0.76, 0.80 and 0.81) by 0.12--0.13 at every size (10 of 10 seeds, $p = 0.002$). Its gap to BPTT (0.93, 0.95 and 0.94) decreases with network size from 0.05 to less than 0.01 (not significant at 200 neurons, $p = 0.68$). E-prop is above chance through the release in its eligibility trace, as in delayed association (Section~\ref{sec-write}): with the presynaptic spike in its place, it remains at chance (0.50 at 100 neurons; 10 of 10 seeds lower, $p = 0.002$).

\section{Discussion}

A synaptic memory is held in a neuron's outgoing synapses but written through its incoming weights: these determine when the neuron spikes, and its spikes set the state of the synapses it sends out. The credit for an incoming weight therefore has to pass through that state. E-prop omits this path and remains at chance on delayed association from the matched initialisation and on delayed XOR, as does BPTT with the path removed on the former; the path is therefore necessary on these tasks. E-prop can still read a synaptic memory when the random initialisation provides a cue preference, but only because its eligibility trace follows the synaptic release: with the presynaptic spike in its place, it remains at chance.

The rule derived here restores the path through the synaptic state forward in time, with one additional trace per synapse, computed from four local variables. Its approximations make the propagation of sensitivities local and leave one non-local quantity beyond those of e-prop, the one-hop feedback weight $b_{j}$. This weight can be reduced to its sign and a gain fixed before training, and the sign can be taken from the neuron's random feedback weight $B_{j}$: during learning, the vector of the $b_{j}$ aligns with $B$ (Section~\ref{sec-needs}). The result, STPx e-prop, requires no weight transport. The locality of STPx e-prop does not reduce its accuracy: it matches STPx e-prop + $b_{j} \delta$ on delayed association and learns delayed XOR with facilitation in every neuron, where the latter fails. It approaches BPTT on all three tasks, with its largest gap on evidence accumulation at 50 neurons (0.05).

\emph{Learning through synaptic memory.} Models trained with BPTT readily store task information in short-term synaptic states~\cite{masse2019-circuit,kozachkov2022-robust}. BPTT writes these memories through the terminal-state path (Section~\ref{sec-write}), so such models assume that a circuit can assign credit through its synaptic states. Experiments show that short-term dynamics shape the induction of long-term plasticity~\cite{froemke2006-contribution} and that long-term plasticity in turn alters short-term dynamics~\cite{markram1996-redistribution,monday2018-long}, but, to our knowledge, there is no direct evidence that long-term plasticity uses the short-term synaptic state as a credit signal. Our results show that it is possible in principle. The credit that BPTT sends along this path can be approximated forward in time from quantities that belong to the neuron itself: the input arriving at its incoming synapses, its spikes and the state of its own terminals, together with one random feedback weight and a broadcast error. Our work identifies the information biological circuits would need to assign credit through short-term synaptic memory. A key challenge is communicating the states of outgoing presynaptic terminals to the incoming synapses where weights are modified. Although rapid axon-to-soma signalling mechanisms exist~\cite{trigo2019-antidromic}, whether they can convey the required information remains unknown. An alternative is postsynaptic memory mediated by back-propagating action potentials, which generate transient dendritic calcium signals~\cite{koester1998-calcium,jones2017-dendritic} that can suppress subsequent excitatory postsynaptic responses~\cite{jones2017-dendritic}. Whether such mechanisms can replace presynaptic STP in memory storage and credit assignment remains an interesting direction for future work.

\emph{Limitations.} The tasks are synthetic working-memory tasks with a single binary decision, and the networks comprise 20--200 neurons, which corresponds to the size of current on-chip recurrent learning~\cite{rostami2022-prop,frenkel2022-reckon} but not to large benchmarks. All tasks have a fixed delay, and each neuron has a single STP type, whereas in cortex the STP of a synapse can depend on its target~\cite{markram1998-differential}. The fixed gain must be of the order of the final magnitude of the feedback weights ($16 \rho_{0}$ here; Fig.~\ref{fig-da}c): a gain four times larger lowers the accuracy by 0.11 (Section~\ref{sec-needs}). It is computed once before training. We measured the sign agreement of $b_{j}$ and $B_{j}$ on delayed association only.

Adaptive neurons provide a memory that e-prop already exploits~\cite{Bellec2020-yh}, and they set a high reference. With spike-frequency adaptation in half of the neurons and static synapses, e-prop reaches 0.95--1.00 on delayed association at every network size (0.97--1.00 with symmetric feedback). Networks with facilitating synapses reach this level only from 100 neurons, also under BPTT (Fig.~\ref{fig-size}), so the difference lies in the network and decreases with its size; at 100 and 200 neurons, STPx e-prop is within 0.04 of e-prop with spike-frequency adaptation (0.95--0.97, compared with 0.98--1.00). On evidence accumulation, compared at 50 neurons only, adaptation is further ahead (0.98 with symmetric e-prop, compared with 0.89 for STPx e-prop and 0.93 for BPTT with facilitation). On delayed XOR the two are comparable (0.96 for STPx e-prop, 0.94--0.95 for e-prop with adaptive neurons; $p = 0.98$; Section~\ref{sec-xor}). A memory in the state of the neuron itself requires no additional credit path; the question addressed here is not whether a memory can be held, but whether a local rule can exploit the state that synapses maintain in any case. Although STP-based networks performed less well than adaptation-based networks on two of the three tasks, their successful learning demonstrates that synaptic dynamics provide an alternative substrate for task-relevant memory and temporal credit assignment, consistent with the broader concept of functional degeneracy in neural systems~\cite{edelman2001-degeneracy}.

\emph{Future work.} The rule extends to neurons with several presynaptic STP states, for example one per class of target, at a proportional cost in memory. Ongoing work tests the generality of the rule in such networks, across unpredictable delays and on more difficult tasks. Finally, the rule keeps the structure of e-prop, which runs on several neuromorphic chips~\cite{rostami2022-prop,frenkel2022-reckon}: it is computed forward in time, stores no trial history, and updates each weight from a learning signal and local traces. Its main obstacle on such hardware is memory: the rule stores five variables per synapse, whereas neuromorphic chips that implement e-prop store its traces per neuron~\cite{frenkel2022-reckon}; a one-variable approximation of the STP model, with a scalar terminal-state trace, would lower this cost.

\section*{Acknowledgement}
The study was funded by the Boehringer Ingelheim Stiftung. We gratefully acknowledge the support of the Institute for Quantitative and Computational Biosciences (IQCB), Johannes Gutenberg University, Mainz, Germany.

\bibliographystyle{IEEEtran}
\bibliography{references}

\end{document}